\documentclass[prl,twocolumn,secnumarabic,amssymb]{revtex4}
\usepackage{graphicx}
\input epsf

\newcommand{\hx}{\hat x}

\begin{document}
 
\title{Stochastic Acceleration in Turbulent Electric Fields Generated 
by 3-D Reconnection}
\author{Marco Onofri, Heinz Isliker, and Loukas Vlahos}
\affiliation{Department of Physics, University of Thessaloniki,
54124 Thessaloniki, Greece}

\begin{abstract}
Electron and proton acceleration in three-dimensional electric and
magnetic fields is studied through test particle simulations. The
fields are obtained by a three-dimensional magnetohydrodynamic
simulation of magnetic reconnection in slab geometry. The
nonlinear evolution of the system is characterized by the growth
of many unstable modes and the initial current sheet is fragmented
with formation of small scale structures.  We inject at random
points inside the evolving current sheet a Maxwellian distribution
of particles. In relatively short time (less than a millisecond)
the particles develop a power law tail.
The acceleration is
extremely efficient and the electrons absorb a large percentage of
the available energy in  a small fraction of the characteristic
time of the MHD simulation, suggesting that resistive MHD codes,
used extensively in the current literature,
are unable to represent the full extent of particle acceleration
in 3D reconnection.
\end{abstract}
\maketitle

It is widely accepted that magnetic reconnection plays a
significant role in converting magnetic energy to thermal energy
and kinetic energy of electrons and protons in laboratory plasmas,
the Earth's magnetosphere,
 the solar corona, and in
extragalactic jets \cite{biskamp,priest2}.

In resistive magnetohydrodynamic models, resistivity breaks the
frozen-in law in a boundary layer, allowing reconnection to occur.
A current sheet can  be spontaneously unstable to resistive
instabilities, like the tearing modes,
which lead to magnetic reconnection \cite{furth,bulanov}.
Many numerical codes have been developed to study the nonlinear
evolution of tearing modes in two-dimensional approximations
\cite{malara}. However, three-dimensional effects may
become important in modifying the spatial structure of the current
sheets and the reconnection rate \cite{dahlburg2002, onofri}.

Different numerical studies have been performed to
investigate collisionless magnetic reconnection using  fluid models,
where magnetic reconnection is made
possible by electron inertia, and kinetic simulations,
in two and three dimensional configurations \cite{pegoraro,buchner1}.
It has been shown that in the three-dimensional kinetic
reconnection the characteristic time scale of the instability is
much faster than that of the two-dimensional tearing mode instability.

The change of the topology of the magnetic field due to magnetic reconnection
allows the release of magnetic energy, which can be responsible for
the acceleration of particles.
In two-dimensional reconnection configurations particle acceleration
has been extensively studied both analytically and numerically
\cite{sonn,cowley}.
The acceleration is caused by the motion of particles along the
electric field in the current sheet, but the magnetic field
 plays a significant role since it influences the trajectory
and therefore the energy gain of the particles. Recently, it has
become clear that it is essential to include in the model the
longitudinal component of the magnetic field, which is parallel to
the electric field in the current sheet \cite{somov}-\cite{lu}.
Studies of particle acceleration with a longitudinal magnetic field component
have also been performed for the case of a magnetic $x$-line
\cite{hamilton}. Moreover, particle acceleration has been studied
through numerical simulations in the framework of two dimensional
resistive and collisionless reconnection
\cite{kliem}-\cite{drake}.

All these studies have been performed for simple electric and magnetic field
configurations, with homogeneous electric fields,  
the investigation of
more complex three-dimensional configurations has though 
shown that more realistic
simulations are necessary to understand particle acceleration in
reconnection regions \cite{birn}.
Test particle simulations  in three-dimensional
electric and magnetic fields, obtained by magnetohydrodynamic simulations of
magnetic reconnection have
shown that idealized
analytical two-dimensional treatments are  too simplified models,
which cannot give a complete understanding of the
problem \cite{shopper,nodes}.

In the present article, we focus our attention on a single current sheet 
and its fragmentation, as yielded by
a resistive magnetohydrodynamics 
simulation of three-dimensional
magnetic reconnection. The initial equilibrium magnetic field
includes a longitudinal component, parallel to the
current sheet, as this was recognized to be an important characteristic
to understand the acceleration of particles. We study the acceleration of
electrons and ions in the electromagnetic fields that result from the 
evolution of this
current sheet and discuss the limitations of the MHD approach 
to reconnecting plasmas and the importance of kinetic effects.

We numerically solve the incompressible, dissipative, magnetohydrodynamics
(MHD) equations in dimensionless units
in a three-dimensional Cartesian domain,
with kinetic and magnetic Reynolds numbers $R_v=5000$ and $R_M=5000$.
We set up the initial condition in such a way to have a plasma that is at rest,
in the frame of reference of our computational domain, permeated by an
equilibrium magnetic field ${\bf B}_0$, sheared along the $\hx$ direction,
with a current sheet in the middle of the simulation domain:
\begin{eqnarray}
{\bf B}_0=B_{y0}{\hat y}+B_{z0}(x){\hat z},
\end{eqnarray}
where $B_{y0}$ is a constant value, which has been set to 0.5 and $B_{z0}$ is given by
\begin{eqnarray}
B_{z0}(x)=\tanh\Big(\frac{x}{0.1}\Big)-\frac{x/0.1}{\cosh^2\Big(\frac{1}{0.1}\Big)}
\end{eqnarray}
In the $y$ and $z$ directions, 
the equilibrium magnetic field is uniform and 
we impose periodic boundary conditions, since we do not expect any
important boundary effects on the development of turbulence.  In 
the inhomogeneous $x$ direction, we impose fixed boundary conditions.
We perturb these equilibrium fields with three-dimensional divergenceless
fluctuations.

\begin{figure}
\epsfxsize=2truein
\centerline{\epsffile{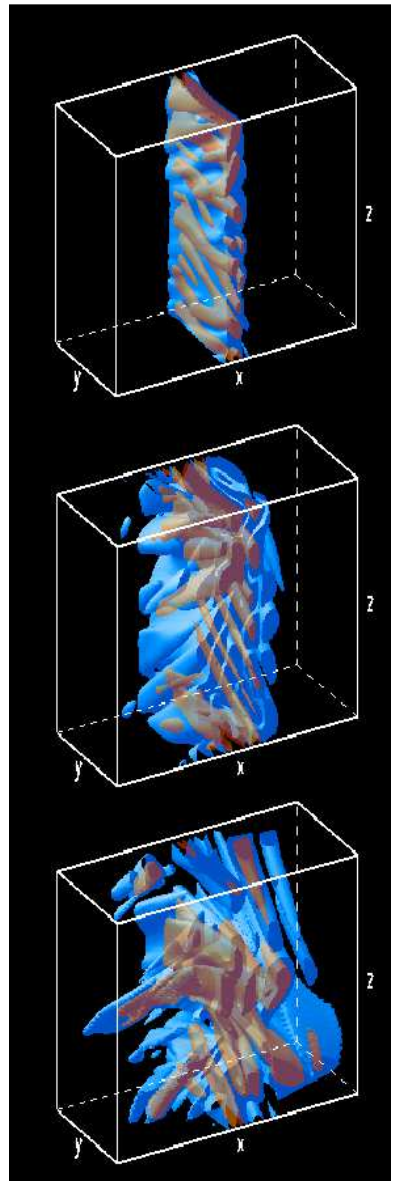}}
\caption{Electric field isosurfaces at $t=50\tau_A$,
$t=200\tau_A$ and $t=400\tau_A$.}
\label{figure1}
\end{figure}

The nonlinear evolution of the system is characterized by the formation
 of small scale structures, especially on the lateral
regions of the computational domain,
and coalescence of magnetic islands in the center.
This behavior is reflected in
the three-dimensional structure of the electric field,
which shows that
the initial equilibrium is destroyed by the formation of current filaments.
For details about this MHD simulation see \cite{onofri}.
After about $t=50 \tau_A$ (where $\tau_A$ is the Alfv\'en time)
the current sheet starts to be fragmented, as can be seen
in Fig.\ \ref{figure1}, where we show the configuration of the electric field
${\bf E}=\eta{\bf J}-{\bf v}\times{\bf B}$ calculated from the MHD simulation.
Figure \ref{figure1} shows the isosurfaces of the electric field at
different times calculated for two different values of the
electric field: the red surface represents higher values and the
blue surface represents lower values.
The structure of the electric field is characterized by small regions of space
where the  field is stronger, surrounded by a larger volume occupied
by lower electric field values.
At later times the fragmentation is more evident,
and at $t=400\tau_A$,  the initial current sheet has been
completely destroyed and the
electric field is highly fragmented.
The strong electric field regions are
acceleration sites for the particles and their distribution in space fills
a larger portion of the simulation box at later times, with increasing
possibility to accelerate a higher number of particles.

To give a measure of the fragmentation of the electric field, we calculated
the fractal dimensions of the fields shown in Fig.\ \ref{figure1},
using the box counting definition of the fractal dimension.
The thresholds are the same that have been used to draw the isosurfaces
shown in Fig.\ \ref{figure1}.
For the fields represented by the blue surfaces in Fig.\ \ref{figure1},
we found fractal dimensions $d=2$, $d=2.5$,  and $d=2.7$
at $t=50\tau_A$, $t=200\tau_A$,  and $t=400\tau_A$,  respectively.
For the more intense electric fields (red surfaces in Fig.\ \ref{figure1}),
the fractal dimensions are $d=1.8$, $d=2$,  and $d=2.4$
at $t=50\tau_A$, $t=200\tau_A$, and $t=400\tau_A$, respectively.
These fractal dimensions
can be considered as a way to quantify the degree of fragmentation of the electric
field and the fraction of space it fills as it evolves in time.

\begin{figure}
\epsfxsize=7cm
\centerline{\epsffile{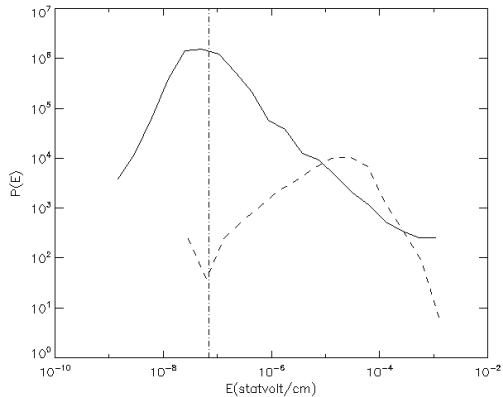}}
\caption{Distribution function of the resistive (dashed line) and convective (solid line)
electric field at $t=50\tau_A$. The vertical line represents the
value of the Dreicer field in the solar corona.}
\label{figure4}
\end{figure}

We calculate the
magnitude $E$ of the electric field at each gridpoint of the simulation domain
and construct the distribution function of these quantities, which is
shown
in Fig.\ \ref{figure4} for $t=50\tau_A $.
We separately plot the resistive and the convective component of the electric field.
The resistive part is less intense than
the convective part, but it is much more important in accelerating particles,
as we verified by performing some simulations where only one of the
two components was used. In the simulations described in this
article
both components of the electric field have been included.

Protons and electrons are injected into the simulation box where they move
under the action of the electric and magnetic
fields, which do not evolve
during the particle motion. This is justified by the fact that
the evolution of the fields is
much slower than the acceleration process, electrons and ions
are accelerated
on a short time scale to very high energies, and in such short times
the fields would not change significantly according to the MHD simulation.

The trajectories of the test particles inside the box are calculated by
solving the relativistic equations of motions,
using a fourth order
Runge Kutta adaptive step-size scheme. Since the magnetic and the electric
fields are given only at a discrete set of points (the grid-points of the
MHD simulation), both fields are interpolated with local three-dimensional
linear interpolation to provide the field values in between grid-points.

The configuration studied here can be applied to many astrophysical situations 
where magnetic reconnection is produced by the development of resistive
instabilities in a current sheet. 
Particularly 
detailed information on the accelerated particles 
is contained  
in the recent data from 
the Ramaty High Energy Solar Spectroscopic Imager (RHESSI) satellite  
obtained from solar flares
\cite{lin}, 
which 
allow a detailed comparison 
with models for particle acceleration.
We thus have chosen  
physical  values in the simulations that are relevant for the solar corona:
$B_0=100 G$, $n_0=10^9 cm^{-3}$, and $l_x=10^9 cm$. 
The corresponding value of the resistivity is $\eta\simeq2\times10^{-6}s$. 
The Dreicer field, which is 
represented by the vertical line in Fig.\ \ref{figure4},
is $E_D\simeq7\times10^{-8} statvolt/cm$.
In each case 50000  test particles (protons or electrons) are
injected in the simulation box with random initial positions and
the three components of the particle initial velocities are randomly
drawn from a Gaussian distribution with a temperature of $1.16\times10^6$~K. 
The electric and magnetic fields used to
accelerate the particles are obtained from the MHD simulation at
$t=50\tau_A$, which corresponds to $t\simeq72$~s. 
Most of the resistive and all of the convective electric fields are
super-Dreicer.
The total available magnetic energy inside the simulation
volume is $W_B=\int(B_0^2/8\pi)dV\sim 10^{32} erg$. 
The particles are followed for
a time $t_p$, much shorter than the time scale 
on which the magnetic configuration evolves. 
The energy distribution is calculated from
the kinetic energy at the time $t_p$ for the particles that are
still in the simulation box and from the energy at the time they exit 
the simulation box for the particles that escaped.

\begin{figure}
\epsfxsize=7cm
\centerline{\epsffile{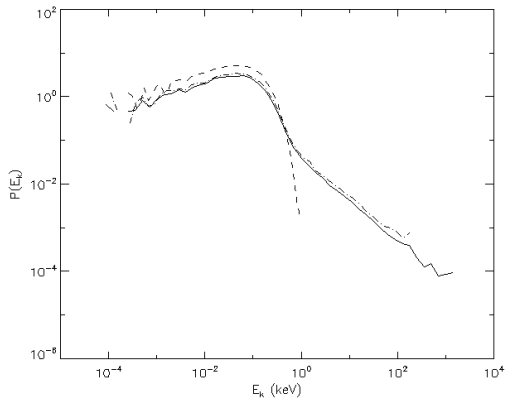}}
\caption{Distribution function of electron kinetic energy at
$t=8\times10^{-5} s$ (solid line), $t=3\times 10^{-5} s$ (dotted-dashed line)
and the initial distribution (dashed line). The electromagnetic
field is given at $t=72$~s.}
\label{figure9}
\end{figure}

For the case of electrons, the particles' energy distribution at
different times is shown in Fig.\ \ref{figure9}. 
Some of the test particles are quickly
accelerated to high energies so that the initial Maxwellian
distribution changes, developing a tail that grows in time. 
The kinetic energy of the electrons increases very rapidly, and
in a short time it equals the  
energy contained in the magnetic field.
Since there is no back reaction of the particles onto the fields, there is no
limit to the energy they can gain until they leave the simulation box.
For this reason we follow the particle motion only as long as their energy is
still less than $50\%$ of the magnetic field energy $W_B$, which is 
up to $t_{pe}=8\times10^{-5}s$.
The maximum kinetic energy at the end of the run turns then out to be
about $1~MeV$. Collisions are not
included in the simulations because the collisional time is about
$t_c=5.5\times10^{-3}$~s, which 
is much longer than $t_{pe}$.
In the final distribution, the logarithmic slope of
the power law tail is $\simeq 1$. 
The power law tails of the
distributions start at an energy of about $1 keV$. The total
number of particles contained in the tail of the distributions
($E_K\geq 1 keV$) is $\simeq6\times10^{37}$ 
for the assumed values of the particle density $n_0$ and length 
$l_x$.  
Below $1 keV$, the electrons have a thermal distribution. 
As we explained in the introduction, it was found in previous works
that substantial acceleration of particles also occurs in a two dimensional 
current sheet. Therefore results similar to those reported
here for $t=50\tau_A$, where the fragmentation of the current sheet has developed,
start to appear at earlier times, but with fewer particles in the 
tail of the distribution.

Turning to protons, we find that acceleration is much less
efficient than for electrons, only at $t_{pi}=3\times10^{-3}s$
they reach a maximum kinetic energy of about $1~MeV$,
with energy distributions that are similar to those of
the electrons. 
Because of the much slower 
acceleration time scale of the protons, the time limit for our
simulation is determined by the electrons, $t_{pe}$ 
($t_{pe}<<t_{pi}$) ---
at times as large as $t_{pi}$, the electrons would
have absorbed all the available magnetic energy $W_B$. 
At the time limit $t_{pe}$ then,
the distribution of the ions has remained close to the initial Maxwellian, 
with just minor gain in enery. 

Since solar flares are closely associated with
magnetic reconnection, the data obtained by the RHESSI satellite  
can be used to validate
theoretical ideas related to the acceleration of particles during
magnetic reconnection  \cite{priest2}. The main constraints
reported from the current RHESSI data are: 1) the energy distribution,
 inferred from the
analysis of the Hard X-rays, is thermal up to an energy of
10-15$\,KeV$ and is smoothly continued  
to a power law
distribution with index around $1$ to $5$ at higher energies; 
2) the
accelerated electrons can reach a maximum energy of
1$\,Mev$, and the ions a maximum energy of
several hundreds of $MeV$; 3) the high energy particles
absorb more than $50\%$ of the  energy released in a solar flare
\cite{lin}, \cite{kontar}.

The results of our simulations show that a decayed and fragmented
current sheet can be a very efficient accelerator. 
The particles absorb a large amount of energy from the
magnetic field in a short time, and the magnetic and electric fields
loose a large fraction of their energy. The back-reaction 
of the particles is though not taken into
account in the test particle simulations reported here, 
which in this sense are not self-consistent. 
Our results suggest that the lifetime of a current sheet 
of this size in the solar corona is very short since energetic
particles absorb a large fraction of the available magnetic energy.
As a consequence of the back-reaction, 
the magnetic and electric field
would change more quickly
than  what the MHD simulation shows,
the acceleration process 
can be expected to be slower, and 
the resulting energy distributions could probably be different. 

The limitations of the presented approach, in the case of the electrons, 
are reflected in the problems of
our results to reproduce all the characteristics of the distributions 
 that are observed in solar flares
(e.g.\ variation of power-law slopes).
In the case of the ions, the situation is different.
Their maximum energy at the time-limit of our simulations
is lower than the energy
that protons usually reach during solar flares. 
We can thus conclude that ions are
not accelerated to the high energies observed during solar flares
by single, isolated, turbulent current sheets.
Stressed and complex large scale magnetic topologies can though form
simultaneously many current sheets \cite{galsgaard}, and 
it has been shown that the interaction of the
ions (and electrons) with many current sheets can be a very
efficient accelerator \cite{turkmani,vlahos}.

Ideal and resistive MHD codes, used widely in many astrophysical
applications, reach the limits of their applicability as soon as
current sheets start to form.  Then the acceleration of electrons will 
start to dominate
the evolution of the current sheet, and 
kinetic phenomena become important.

The fast acceleration that we observe in the simulation is also
due to the high (anomalous) resistivity that we use and the
consequent high values of the parallel electric field. Such a value
of the resistivity is necessary, from a practical point of view,
in order to simulate the process of reconnection in reasonable CPU time, and,
in astrophysical terms, to
release the observed magnetic energy on a time scale comparable to the flare
process. Anomalous resistivity is  realistic only in limited
regions (where thin current sheets are formed) in the solar
corona. Classical resistivity, which is extremely small in the
corona, is appropriate for the slow evolution of the large scale
magnetic structures prior to the formation of the current sheets.

In summary, if large scale 3D current sheets are formed in the solar
corona, they evolve slowly in the beginning and soon collapse, 
forming a turbulent super-Dreicer
electric field environment.  These fragmented current sheets are
very efficient electron accelerators and mark the end of the
applicability of MHD. MHD is not a good approximation to describe
the entire process of reconnection in the range of parameters that
is of interest for fast magnetic dissipation and particle
acceleration, and a self-consistent kinetic treatment becomes
necessary.

\end{document}